\documentclass{ifacconf}

\input{doc/header}

\newcommand{\arxivTF}[2]{#1}

\begin{document}
\begin{frontmatter}

\title{Robust Performance Analysis for Time-Varying Multi-Agent Systems with Stochastic Packet Loss\textsuperscript{1}}

\author[First]{Christian Hespe}
\author[First]{Herbert Werner}

\address[First]{Hamburg University of Technology, Institute of Control Systems, 21073 Hamburg, Germany. \{christian.hespe, h.werner\}@tuhh.de}

\thanks{© 2023 the authors. This work has been accepted to IFAC for publication under a Creative Commons Licence CC-BY-NC-ND}

\begin{abstract}
Recently, a scalable approach to system analysis and controller synthesis for homogeneous multi-agent systems with Bernoulli distributed packet loss has been proposed.
As a key result of that line of work, it was shown how to obtain upper bounds on the \(H_2\)-norm that are robust with respect to uncertain interconnection topologies.
The main contribution of the current paper is to show that the same upper bounds hold not only for uncertain but also time-varying topologies that are superimposed with the stochastic packet loss.
Because the results are formulated in terms of linear matrix inequalities that are independent of the number of agents, multi-agent systems of any size can be analysed efficiently.
The applicability of the approach is demonstrated on a numerical first-order consensus example, on which the obtained upper bounds are compared to estimates from Monte-Carlo simulations.

\end{abstract}


\end{frontmatter}

\section{Introduction}\label{sec:intro}
In many cases, approaching large-scale control problems in a distributed or decentralized manner is desirable if not unavoidable.
Compared to centralized approaches, the aforementioned ones allow to split a single large-scale problem into a collection of small ones, which are easier to solve, both from engineering and computational standpoint \citep{Massioni2009}.

A particular class of distributed control systems are those in which information is exchanged using some form of communication network.
As outlined by \citet{Schenato2007}, these kinds of networks are inherently unreliable in transmitting information and are thus subject to stochastic packet loss.
Nonetheless, many existing works on distributed control systems neglect this source of stochasticity and consider the information exchange to be ideal and deterministic.
This is notably the case in the decomposable systems framework as originally proposed by \citet{Massioni2009} for linear time-invariant systems.
Only recently, \citet{Hespe2022} proposed an extension of the framework to Markov jump linear systems (MJLSs) allowing to study Bernoulli distributed packet loss with uniform loss probability while preserving the scalability of the approach.
As part of that work, the authors suggest treating the system's interconnection structure as a parametric uncertainty.
Their existing analysis does however not cover time-varying interconnection structures, in contrast to similar results for linear time-invariant systems (see, e.g., \citep{Pilz2009}).
The main result of the current paper is to close this gap and establish that the analysis conditions proposed by \citet{Hespe2022} are sufficient not only for uncertain but also time-varying interconnections.

To capture the effects of the superimposed packet loss and time-varying graphs, standard MJLSs with a single stochastic switching law as described by \citet{Costa2005} are not sufficient.
While the packet loss is stochastic, the switching of the interconnection structure is not, such that an additional deterministic switching law is required.
Such systems have been considered in the literature before, termed as \emph{switched} MJLSs:
For example, their stability properties have been studied by \citet{Bolzern2010, Song2016} and \citet{Qu2018}, while \citet{Zhang2011} and \citet{Hou2013} considered system performance in terms of (stochastic) \(l_2\) and \(l_2-l_\infty\) system norms.
Importantly, we only consider a special kind of switched MJLSs for which the stochastic switching is independent in time.
As outlined for standard MJLSs by \citet{Fioravanti2012}, it is possible to derive simplified analysis conditions for these systems, which has not been exploited in any of the previously mentioned papers on switched MJLSs.

On the other hand, the consensus problem, which the multi-agent system (MAS) framework of this paper contains as a special case, on switching graphs -- covering \emph{deterministic} but not \emph{stochastic} packet loss -- has been studied earlier by, amongst others, \citet{Ren2005} and \citet{Mesbahi2010}.
In contrast to the probabilistic statements made in the current paper, i.e., mean-square stability and mean output energy, papers along that line of work present deterministic results.
However, these approaches require stronger assumption on the connectivity of the underlying graphs, which can be difficult to satisfy in practice.

The main contributions of this paper come in the form of two theorems:
Theorem~\ref{thm:h2} establishes stability and performance bounds for general switched MJLSs with independent stochastic switching and Theorem~\ref{thm:h2_mas} specializes the result to obtain scalable conditions for networked MASs in the form of linear matrix inequalities (LMI).
Both theorems are presented in Section~\ref{sec:analysis}, which is preceded by problem setup and modelling in Section~\ref{sec:problem}.
Section~\ref{sec:demo} demonstrates the application of the results on a first-order consensus example and Section~\ref{sec:conclusions} concludes the paper.

\section{Problem Setup}\label{sec:problem}
\subsection{Notation \& Definitions}\label{sec:problem_notation}
We let \(I_N\) denote the \(N \times N\) identity matrix.
\(M_1 \otimes M_2\) is the Kronecker product of two matrices and \(M \succ (\succeq)\ 0\) or \(M \prec (\preceq)\ 0\) mean that \(M\) is positive or negative (semi-)definite, respectively.
For block-diagonal concatenation, we use the notation \(\diag(M_1, M_2, \ldots)\).
\(\lambda_i(M)\) denotes the \(i\)th eigenvalue of the matrix~\(M\).
The Euclidean vector norm is \(\|x\| \coloneqq \sqrt{x^T x}\) and the same symbol is used to denote the stochastic signal \(l_2\)-norm \(\|z\|^2 \coloneqq \sum_{k = 0}^{\infty} \expect\big[\|z_k\|^2\big]\).

For modelling interconnection structures, we will use graphs \(\mathcal{G} \coloneqq (\mathcal{V}, \mathcal{E})\), composed of the vertex set \(\mathcal{V} = \{\, 1, 2, \ldots, N \,\}\) and the edge set \(\mathcal{E} \subset \mathcal{V} \times \mathcal{V}\).
Defining \(e_{ij} \coloneqq (j, i)\), \(\mathcal{G}\) is called undirected if \(e_{ij} \in \mathcal{E} \Leftrightarrow e_{ji} \in \mathcal{E}\) and graphs are assumed to contain no self loops, i.e., \(e_{ii} \notin \mathcal{E}\).
A sequence of vertices \((v_1, v_2,\ldots, v_n) \in \mathcal{V}^n\) is called path if \((v_i, v_{i+1}) \in \mathcal{E}\) for \(1 \leq i \leq n-1\).
If, for all \(i \in \mathcal{V}\), there exists a path to all other \(j \in \mathcal{V} \setminus \{i\}\), \(\mathcal{G}\) is said to be connected.
Furthermore, the Laplacian is given element-wise as \(L(\mathcal{G}) \coloneqq [l_{ij}(\mathcal{G})]\), where
\begin{equation}\label{eq:laplacian}
    l_{ij}(\mathcal{G}) \coloneqq \begin{cases}
        -1 & if \(i \neq j\) and \(e_{ij} \in \mathcal{E}\), \\
         0 & if \(i \neq j\) and \(e_{ij} \notin \mathcal{E}\), \\
        -\textstyle{\sum_{l \neq i}} l_{il}(\mathcal{G}) & if \(i = j\).
    \end{cases}
\end{equation}

\subsection{Stochastic Packet Loss as Random Graphs}\label{sec:problem_graphs}
The focus of this paper is on analysing the effect of stochastic packet loss on networked MASs whose interconnection structure, i.e., which agents are neighbouring each other and exchanging information, is changing over time.
This is in contrast to earlier work in \citet{Hespe2022}, where the information exchange is affected by stochastic packet loss, but the underlying interconnection structure is assumed to be constant.

First, we describe the uncertainty about the interconnection structure.
As stated above, the neighbouring relations between the agents in the MAS will be modelled using graphs, where there is a one-to-one mapping between the agents and vertices of the graph such that \(N\) corresponds the number of agents in the MAS.
The set of agents -- and thus vertices -- is assumed to be constant.
In the MAS, an edge~\(e_{ij}\) is read as agent~\(i\) receiving information from agent~\(j\).
However, in contrast to the vertex set, the edge set is time-varying, representing the changes in the interconnection structure.
Given \(\mathcal{F} \coloneqq \{\,\mathcal{E}_1, \mathcal{E}_2, \ldots\,\} \subset 2^{\mathcal{V} \times \mathcal{V}}\), the family of all admissible edge sets, and the discrete time index \(k \geq 0\), the time dependency is described by the switching sequence~\(\nu\) with \(\nu_k \in \mathcal{K} \coloneqq \{\, 1, \ldots, |\mathcal{F}| \,\}\) such that \(\mathcal{E}_{\nu_k}\) is the edge set at time~\(k\).
Furthermore, introduce \(\mathcal{G}_j \coloneqq (\mathcal{V}, \mathcal{E}_j)\) and \(\mathcal{E}^0 \coloneqq \bigcup_{\mathcal{E}_j \in \mathcal{F}} \mathcal{E}_j\), the union of all admissible edge sets.
We will then make the following assumption to capture the uncertainty:
\begin{assum}\label{ass:bounded_laplacian}
    There exist constants \(0 < \ubar{\lambda} \leq \bar{\lambda}\) such that
    \begin{equation}\label{eq:admissible_graphs}
        \mathcal{G}_j \in \big\{\,\mathcal{G} \text{ undirected} : \ubar{\lambda} \leq \lambda_i\big(L(\mathcal{G})\big) \leq \bar{\lambda}, 2 \leq i \leq N \,\big\}
    \end{equation}
    for all \(j \in \mathcal{K}\).
\end{assum}
Note that Assumption~\ref{ass:bounded_laplacian} implies that \(\mathcal{G}_{\nu_k}\) is connected for all \(k \geq 0\).
In practice, \(\ubar{\lambda}\) and \(\bar{\lambda}\) can be estimated from the maximum degree of the admissible graphs and Cheeger's inequality \citep[Section 2.4.2]{Mesbahi2010}.

On top of the changing interconnection structure, the communication network of the MAS is subject to stochastic packet loss.
For each \(e_{ij} \in \mathcal{E}^0\), introduce a stochastic process \(\{\alpha_k^{ij}\}\) with \(\alpha_k^{ij} \in \{\,0, 1\,\}\).
A value of \(\alpha_k^{ij} = 1\) is interpreted as \(e_{ij}\) being active at time~\(k\), i.e., information may be transmitted on \(e_{ij}\) if it is in \(\mathcal{E}_{\nu_k}\), while \(\alpha_k^{ij} = 0\) means no information is transmitted from agent~\(j\) to agent~\(i\) regardless of \(\mathcal{E}_{\nu_k}\).
For this paper, we restrict our analysis to Bernoulli distributed packet loss that is independent in time.
Additionally, the loss between two distinct pairs of vertices is assumed to be independent.
This is formalized in the following assumption:
\begin{assum}\label{ass:bernoulli}
    The stochastic processes \(\{\alpha_k^{ij}\}\) are \emph{partially} independent and identically Bernoulli distributed such that, for all \(k, k' \geq 0\), \(e_{ij}, e_{rs} \in \mathcal{E}^0\), we have
    \begin{align}\label{eq:bernoulli_probability}
        \probab\big(\alpha_k^{ij} = 1\big) &= p & \probab\big(\alpha_k^{ij} = 0\big) &= 1 - p
    \end{align}
    with \(p \in [0, 1]\) and \(\alpha_k^{ij}\) and \(\alpha_{k'}^{rs}\) are independent random variables whenever \(k \neq k'\) or \((r, s) \neq (i, j) \land (r, s) \neq (j, i)\).
\end{assum}

The combination of the interconnection structure switching and the packet loss enlarges the set of graphs that may represent the exchange of information.
Instead of \(\mathcal{F}\), the family of admissible edge sets under packet loss is described by
\begin{equation}
    \tilde{\mathcal{F}} \coloneqq \big\{\, \tilde{\mathcal{E}} : \exists \mathcal{E} \in \mathcal{F}, \tilde{\mathcal{E}} \subseteq \mathcal{E} \,\big\} = \big\{\,\tilde{\mathcal{E}}_1, \tilde{\mathcal{E}}_2, \ldots\,\big\}.
\end{equation}
Analogously to \(\mathcal{G}_j\) above, define the switched graphs \(\tilde{\mathcal{G}}_i \coloneqq (\mathcal{V}, \tilde{\mathcal{E}}_i)\).
Note that in contrast to \(\mathcal{G}_j\), the graphs \(\tilde{\mathcal{G}}_i\) do not have to be undirected nor connected.
We can then take a function \(\mu \colon \mathcal{E}^0 \to \{\, 1, \ldots, |\mathcal{E}^0| \,\}\) that assigns a unique integer to each \(e_{ij} \in \mathcal{E}^0\) and define
\begin{equation}\label{eq:stochastic_switching_law}
    \sigma_k \coloneqq 1 + \sum_{e_{ij} \in \mathcal{E}_{\nu_k}} \alpha_k^{ij} 2^{\mu(e_{ij}) - 1}
\end{equation}
such that \(\sigma_k \in \tilde{\mathcal{K}} \coloneqq \{\, 1, \ldots, |\tilde{\mathcal{F}}| \,\}\) and \(\tilde{\mathcal{G}}_{\sigma_k}\) describes the exchange of information in the MAS at time~\(k\).

Because of the random packet loss, \(\{\tilde{\mathcal{G}}_{\sigma_k}\}\) -- and accordingly \(\{L(\tilde{\mathcal{G}}_{\sigma_k})\}\) -- are stochastic processes.
Defining the two shorthand notations \(L_j \coloneqq L(\mathcal{G}_j)\) and \(\tilde{L}_i \coloneqq L(\tilde{\mathcal{G}}_i)\), we have the following lemma:
\begin{lem}\label{lem:expected_laplacians}
    Given a communication network that satisfies Assumptions~\ref{ass:bounded_laplacian} and \ref{ass:bernoulli}, the conditional expectations of the graph Laplacian at time \(k \geq 0\) are given by
    \begin{subequations}\label{eq:expected_laplacians}\begin{align}
        \expect\big[\tilde{L}_{\sigma_k} \big| \nu_k = j\big]                        &= p L_j, \label{eq:expected_laplacians_simple}\\
        \expect\big[\tilde{L}_{\sigma_k}^T \tilde{L}_{\sigma_k} \big| \nu_k = j\big] &= p^2 L_j^2 + 2p(1-p) L_j. \label{eq:expected_laplacians_squared}
    \end{align}\end{subequations}
\end{lem}
\begin{pf}
    By the network model outlined above, information is transmitted via an edge~\(e_{ij}\) at time~\(k\) if \(e_{ij} \in \mathcal{E}_{\nu_k}\) \emph{and} \(\alpha_k^{ij} = 1\).
    For the graph Laplacian, written element-wise as \(\tilde{L}_{\sigma_k} = [\tilde{l}_{ij}(\sigma_k)]\), this results in
    \begin{equation}\label{eq:random_laplacian}
        \tilde{l}_{ij}(\sigma_k) = \begin{cases}
            -\alpha_k^{ij}  & if \(i \neq j\) and \(e_{ij} \in \mathcal{E}_{\nu_k}\), \\
             0              & if \(i \neq j\) and \(e_{ij} \notin \mathcal{E}_{\nu_k}\), \\
            -\textstyle{\sum_{l \neq i}} \tilde{l}_{il}(\sigma_k) & if \(i = j\).
        \end{cases}
    \end{equation}
    By applying Lemma~4 from \citet{Hespe2022}, we obtain the expectations as stated above, where we utilized that all \(\mathcal{G}_j\) are undirected by Assumption~\ref{ass:bounded_laplacian}, which in turn implies that all \(L_j\) are symmetric.
\end{pf}

\subsection{Decomposable Jump Linear Systems}\label{sec:problem_decomposable}
Having introduced the networking model in the previous subsection, we focus on the dynamic model of the MAS.
In order to obtain scalable analysis conditions later, we build upon the decomposable systems framework introduced by \citet{Massioni2009}.
In their terms, a matrix~\(M\) is called decomposable if it can be split into a decoupled part~\(M^d\) and a coupled part~\(M^c\) as \(M = I_N \otimes M^d + S \otimes M^c\), where \(S\) is termed the \emph{pattern matrix}.
Accordingly, a linear time-invariant state-space model is called decomposable if all of its matrices are decomposable with respect to the same pattern matrix.

This concept was extended to MJLSs by \citet{Hespe2022} to obtain scalable performance analysis conditions for MASs subject to stochastic packet loss (see \citep{Costa2005} for a general introduction to MJLSs).
Based on that line of work, we will in this paper consider systems of the form
\begin{equation}\label{eq:jls}
    P: \,\left\{\enspace\begin{aligned}
        \xi_{k+1} &= A_{\sigma_k, \nu_k} \xi_k + B_{\sigma_k, \nu_k} w_k, \\
        z_k       &= C_{\sigma_k, \nu_k} \xi_k + D_{\sigma_k, \nu_k} w_k,
    \end{aligned}\right.
\end{equation}
where \(\nu_k\) and \(\{\sigma_k\}\) are the switching sequences introduced above, \(\xi_k \in \real^{N n_\xi}\) is the dynamic state, and \(w_k \in \real^{N n_w}\) and \(z_k \in \real^{N n_z}\) are the performance in- and output of the system, respectively.
Furthermore, the switched system matrices are given by
\begin{equation}\label{eq:matrix_structure}
    A_{i,j} = I_N \otimes A^d + \tilde{L}_i \otimes A^c + L_j \otimes A^p
\end{equation}
and similarly for \(B_{i,j}\), \(C_{i,j}\) and \(D_{i,j}\).
This specific structure with a deterministic (\(L_j\)) \emph{and} a stochastic (\(\tilde{L}_i\)) pattern matrix is an extension of decomposable MJLSs that was proposed by \citet{Hespe2023} to allow for deterministic performance channels, which are required to prevent the performance output from vanishing in case of small transmission probabilities~\(p\).
It is demonstrated in Section~\ref{sec:demo_consensus} for a first-order consensus example how a network MAS can be brought into the form of the MJLS~\eqref{eq:jls} with matrix structure \eqref{eq:matrix_structure}.

Contrasting the previous works \citep{Hespe2022, Hespe2023}, \(P\) is scheduled by \emph{two} switching sequences to accommodate both the arbitrary switching of the interconnection structure and the stochastic packet loss.
Furthermore, the distribution of \(\{\sigma_k\}\) depends on \(\nu_k\) through \eqref{eq:stochastic_switching_law}.
\(P\) is therefore not a classical but a \emph{switched} MJLS.

\section{Robust Performance Analysis}\label{sec:analysis}
Based on the stochastic system model introduced above, we derive conditions in the form of linear matrix inequalities (LMIs) that can calculate upper bounds on the robust \(H_2\)-norm of the networked MAS in a scalable manner.
First, we derive conditions for the switched MJLS~\eqref{eq:jls}, which we then specialize to the specific MAS considered in this paper.

\subsection{Conditions for \(H_2\)-Analysis}\label{sec:analysis_h2}
Initially, consider the switched MJLS~\eqref{eq:jls} for general system matrices, i.e., without enforcing \eqref{eq:matrix_structure}.
Under Assumption~\ref{ass:bernoulli}, the stochastic processes \(\{\alpha_k^{ij}\}\) are independent in time.
As such, the distribution of \(\sigma_k\) does not depend on \(\sigma_{k-1}\) and is given by
\begin{align}\label{eq:jls_probability}
    \probab\left(\sigma_k = i\,\middle|\, \nu_k = j\right) = t_i^j & &
    \text{with} & &
    \sum_{i \in \tilde{\mathcal{K}}} t_i^j = 1
\end{align}
and \(t_i^j \geq 0\) for all \(j \in \mathcal{K}\).
This is a special type of MJLS which allows for deriving simplified analysis conditions compared to the general case.

Switched MJLS with general \(\{\sigma_k\}\) have been considered in literature before.
In \citet{Bolzern2010}, stability properties of continuous-time switched MJLS for which the distribution of \(\{\sigma_k\}\) is independent of \(\nu_k\) were studied.
An extension to MJLS for which \(\{\sigma_k\}\) depends on \(\nu_k\) was reported by \citet{Song2016}.
\citet{Zhang2011} proposed a bounded real lemma for discrete-time MJLS with a single (stochastic) switching sequence but time-varying distribution of \(\{\sigma_k\}\) and the generalized \(H_2\)-norm was studied for general switched MJLS by \citet{Hou2013}.
However, none of the above approaches considers switching sequences \(\{\sigma_k\}\) that are independent in time.
In the following, we derive analysis conditions for the \(H_2\)-norm that exploit this property to achieve a computational advantage similar to the approach proposed by \citet{Fioravanti2012} for non-switched MJLS.

There exists a variety of different stability concepts for stochastic switched systems, amongst which are almost sure stability, stability in probability and mean-square stability.
In this paper, we focus on the latter because it implies stability as in the former two definitions \citep{Costa2005}.
The following definition was adapted from \citet{Costa2005} to encompass the additional switching sequence~\(\nu\):
\begin{defn}\label{def:ms_stability}
    The switched MJLS \eqref{eq:jls} is \emph{mean-square stable} if
    \begin{equation*}
        \lim\limits_{k \to \infty} \expect\left[\|\xi_k\|^2\right] = 0
    \end{equation*}
    for all initial conditions~\(\xi_0\), initial distributions~\(\sigma_0\) and switching sequences~\(\nu\).
\end{defn}

In addition to studying the stability properties of the switched MJLS, we will furthermore consider the system performance in terms of the \(H_2\) system norm.
For general non-switched MJLS, this norm was considered by \citet{Costa1997}.
As for mean-square stability, the definition used in that paper is adjusted to encompass switched MJLS as follows:
\begin{defn}\label{def:h2_norm}
    The \(H_2\)-norm of the switched MJLS~\eqref{eq:jls} is
    \begin{equation}\label{eq:h2_norm}
        \|P\|_{H_2}^2 \coloneqq \sup_{\nu} \sum_{s = 1}^{N n_w} \big\|z^{\nu,s}\big\|^2,
    \end{equation}
    where \(z^{\nu,s}\) is the output obtained from applying a discrete unit-impulse to the \(s\)th input with switching sequence~\(\nu\) and \(\xi_0 = 0\).
\end{defn}

Having defined the \(H_2\)-norm for switched MJLS, we can then state a set of LMI conditions to guarantee mean-square stability as well as an upper bound on the system performance in terms of the norm:
\begin{thm}\label{thm:h2}
    The switched MJLS~\eqref{eq:jls} with switching probability~\eqref{eq:jls_probability} is mean-square stable and \(\|P\|_{H_2} < \gamma\) if there exist \(Q \succ 0\) and a symmetric \(Z\) with \(\trace(Z) < \gamma^2\) such that
    \begin{subequations}\label{eq:h2_lmi}\begin{align}
        \sum_{i \in \tilde{\mathcal{K}}} t_i^j \big(A_{i,j}^T Q A_{i,j} + C_{i,j}^T C_{i,j}\big) &\prec Q, \label{eq:h2_lmi_gramian}\\
        \sum_{i \in \tilde{\mathcal{K}}} t_i^j \big(B_{i,j}^T Q B_{i,j} + D_{i,j}^T D_{i,j}\big) &\prec Z  \label{eq:h2_lmi_trace}
    \end{align}\end{subequations}
    hold for all \(j \in \mathcal{K}\).
\end{thm}
\begin{pf}
    \arxivTF{See Appendix~\ref{app:h2_proof}}{See \citep{ExtendedVersion}.}
\end{pf}

As in the conditions proposed by \citet{Fioravanti2012}, the performance test in Theorem~\ref{thm:h2} consist of two matrix variables independent of the cardinality of \(\mathcal{K}\) and \(\tilde{\mathcal{K}}\).
However, it is necessary to check the two LMIs~\eqref{eq:h2_lmi} for each \(j \in \mathcal{K}\), which can lead to tractability issues if the set~\(\mathcal{K}\) is large.
Nonetheless, this results in a sizeable reduction in computational complexity compared to analysis conditions for MJLSs without independence in time of \(\{\sigma_k\}\), since those require to evaluate LMIs for all \(j \in \mathcal{K}\) \emph{and} \(i \in \tilde{\mathcal{K}}\) \citep{Costa2005, Qu2018}.

\subsection{Application to Multi-Agent Systems}\label{sec:analysis_mas}
As the second step, we specialize the conditions for general switched MJLSs with switching probability~\eqref{eq:jls_probability} to decomposable MJLSs with system matrix structure~\eqref{eq:matrix_structure}.
Because of the specific structure of the system matrices, we can apply Lemma~\ref{lem:expected_laplacians} to analytically calculate the conditional expectations in \eqref{eq:h2_lmi}, removing the need to iterate all \(i \in \tilde{\mathcal{K}}\).
Furthermore, by relying on Assumption~\ref{ass:bounded_laplacian} and convexity of the LMI conditions with respect to the eigenvalues of the Laplacian, we can formulate the analysis conditions independent of the number of agents.
Similar conditions were proposed by \citet[Lemma 9]{Hespe2022}, the previous result does however not cover the case of switching interconnection structures and is valid only for uncertain but constant \(\nu\).

\begin{thm}\label{thm:h2_mas}
    The switched MAS~\eqref{eq:jls} satisfying Assumptions~\ref{ass:bounded_laplacian} and \ref{ass:bernoulli} is mean-square stable and \(\|P\|_{H_2}^2 < \beta^2 + (N-1) \gamma^2\) if there exist \(Q \succ 0\) and symmetric \(Z_1\), \(Z_2\) with \(\trace(Z_1) < \gamma^2\) and \(\trace(Z_2) < \beta^2\) such that
    \begin{subequations}\label{eq:h2_mas_lmi}\begin{align}
        \bar{A}_\lambda^T Q \bar{A}_\lambda + \bar{C}_\lambda^T \bar{C}_\lambda + \bar{p}_\lambda \big(A^{cT} Q A^c + C^{cT} C^c\big) &\prec Q, \label{eq:h2_mas_lmi_gramian}\\
        \bar{B}_\lambda^T Q \bar{B}_\lambda + \bar{D}_\lambda^T \bar{D}_\lambda + \bar{p}_\lambda \big(B^{cT} Q B^c + D^{cT} D^c\big) &\prec Z_1, \label{eq:h2_mas_lmi_trace} \\
        A^{dT} Q A^d + C^{dT} C^d &\prec Q, \label{eq:h2_mas_lmi_asymptotic_gramian}\\
        B^{dT} Q B^d + D^{dT} D^d &\prec Z_2, \label{eq:h2_mas_lmi_asymptotic_trace}
    \end{align}\end{subequations}
    hold for \(\lambda \in \{\,\ubar{\lambda}, \bar{\lambda}\,\}\), where \(\bar{p}_\lambda \coloneqq 2p(1-p)\lambda\) and \(\bar{A}_\lambda \coloneqq A^d + \lambda (pA^c + A^p)\) and equally for \(\bar{B}_\lambda\), \(\bar{C}_\lambda\) and \(\bar{D}_\lambda\).
\end{thm}
\begin{pf}
    \arxivTF{See Appendix~\ref{app:h2_mas_proof}}{See \citep{ExtendedVersion}.}
\end{pf}
\begin{rem}\label{rem:scalability}
    It is important to note that the size of the variables is drastically reduced.
    While \(Q \in \real^{n_x \times n_x}\) and \(Z_1, Z_2 \in \real^{n_w \times n_w}\) in Theorem~\ref{thm:h2_mas}, Theorem~\ref{thm:h2} requires \(Q \in \real^{N n_x \times N n_x}\) and \(Z \in \real^{N n_w \times N n_w}\).
\end{rem}
\begin{rem}\label{rem:synthesis}
    Although the results presented here are only concerned with system analysis, the conditions in \eqref{eq:h2_mas_lmi} can be readily extended to robust controller synthesis along the lines of \citep{Hespe2023}.
\end{rem}
\begin{rem}\label{rem:consensus}
    The LMI~\eqref{eq:h2_mas_lmi_asymptotic_gramian} is only feasible if all eigenvalues of \(A^d\) are strictly inside the unit circle.
    However, for many relevant problems, e.g., formation control, \(A^d\) will have a spectral radius of 1 (see also the example in Section~\ref{sec:demo_consensus}).
    As discussed by \citet{Hespe2022}, it is in that case still possible to analyse the systems by projecting it onto the disagreement space, i.e., only considering the system relative to its centre of mass.
    It is then sufficient to check \eqref{eq:h2_mas_lmi_gramian} and \eqref{eq:h2_mas_lmi_trace}.
    This is akin to neglecting the 0 eigenvalue of the Laplacians, a common technique when analysing MASs of linear time-invariant agents \citep{Fax2004}.
\end{rem}

\section{Numerical Demonstration}\label{sec:demo}
To study the properties of the upper bound on the \(H_2\)-norm that can be obtained from Theorem~\ref{thm:h2_mas}, we will in this section demonstrate its application.
The LMIs from the theorem are implemented in \textsc{Matlab} using \textsc{Yalmip} \citep{Loefberg2004} and the source code that generates the figures is available at \citep{Sourcecode}.

\subsection{First-Order Consensus with Packet Loss}\label{sec:demo_consensus}
The example we consider for studying Theorem~\ref{thm:h2_mas} is that of first-order consensus with packet loss and time-varying interconnection structure.
Using \(x_k^i, u_k^i, w_k^i \in \real\) as the state, input and disturbance of agent~\(i\), respectively, the dynamics of the agents are given by
\begin{equation}
    x_{k+1}^i = x_k^i + u_k^i + w_k^i. \label{eq:consensus_agents}
\end{equation}
For each agent, the input \(u_k^i\) is determined by exchanging information in a local neighbourhood.
In the current scenario, we consider this neighbourhood to be time-varying for two reasons:
\begin{enumerate}[label=\roman*)]
    \item the interconnection topology is switching arbitrarily amongst graphs that adhere to Assumption~\ref{ass:bounded_laplacian}, and
    \item the transmitted information is randomly lost as described by Assumption~\ref{ass:bernoulli}.
\end{enumerate}
Overall, this leads to the \emph{consensus protocol}
\begin{equation}
    u_k^i = \kappa \sum_{\mathclap{(j,i) \in \mathcal{E}_{\nu_k}}} \alpha_k^{ij} \big(x_k^j - x_k^i\big), \label{eq:consensus_protocol}
\end{equation}
where \(\kappa > 0\) is a gain parameter of the protocol.
Details on the consensus problem with time-invariant interconnection structure and without packet loss can be found in \citep{OlfatiSaber2007}.

Stacking up the states as \(x_k^T = [x_k^1, x_k^2, \ldots x_k^N]\), and analogously for the input~\(u_k\) and disturbance~\(w_k\), the consensus problem can be brought into the form of the switched MAS~\eqref{eq:jls} with
\begin{align*}
    A^d &=  1,      & B^d &= 1, & C^d &= 0, & D^d &= 0, \\
    A^c &= -\kappa, & B^c &= 0, & C^c &= 0, & D^c &= 0, \\
    A^p &=  0,      & B^p &= 0, & C^p &= 1, & D^p &= 0,
\end{align*}
where we use the consensus error \(z_k = L_{\nu_k} x_k\) as the performance output.
Note that \(z_k\) is calculated from the deterministic Laplacian \(L_{\nu_k}\), not the stochastic \(\tilde{L}_{\sigma_k}\), such that a deviation between two agents is penalized even if the packets transmitted between them are lost.
This ensures that the consensus error is accurately reflected in the performance measure even for small transmission probabilities \(p \to 0\).

In the following, we will consider only \(N = 20\) and \(\kappa = 0.1\).
Note that, in line with Remark~\ref{rem:consensus}, only the LMIs~\eqref{eq:h2_mas_lmi_gramian} and \eqref{eq:h2_mas_lmi_trace} are checked for Theorem~\ref{thm:h2_mas} because the spectral radius of \(A^d\) is 1.

\subsection{Influence of \(\ubar{\lambda}\) and \(\bar{\lambda}\) on the \(H_2\)-Norm Bound}\label{sec:demo_interval}
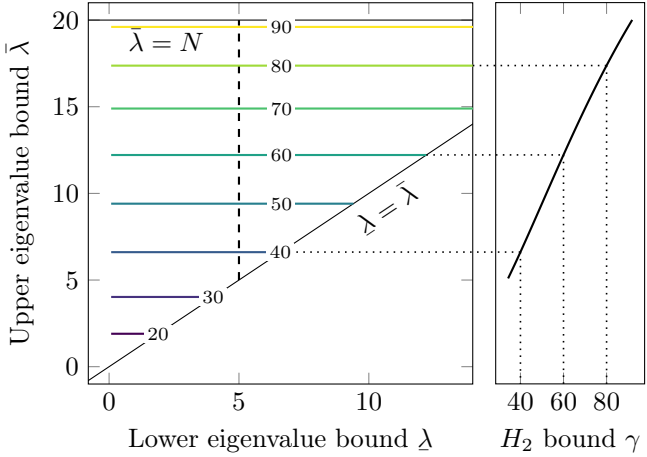
\begin{figure}
    \centering
    \begin{tikzpicture}   
    \pgfplotstableread[col sep=comma]{figures/data/eigenvalue_contour.csv} \contourtable
    
    \pgfmathsetmacro{\i}{0}
    
    \pgfplotsforeachungrouped \x in {1,2,...,8}{
        \pgfplotstablegetelem{\i}{0}\of{\contourtable}
        \pgfmathsetmacro{\level}{\pgfplotsretval}
        \ifdefined\levels
            \edef\levels{\levels, \level}
        \else
            \edef\levels{\level}
        \fi
        
        \pgfmathtruncatemacro{\nextrow}{\i+1}
        \pgfplotstablegetelem{\nextrow}{1}\of{\contourtable}
        \pgfmathsetmacro{\coord}{\pgfplotsretval}
        \ifdefined\coords
            \edef\coords{\coords, \coord}
        \else
            \edef\coords{\coord}
        \fi
        
        \pgfplotstablegetelem{\i}{1}\of{\contourtable}
        \pgfmathtruncatemacro{\i}{1+\i+\pgfplotsretval}
    }
    
    \pgfplotstableset{
        create on use/level/.style={create col/set list/.expanded={\levels}},
        create on use/coord/.style={create col/set list/.expanded={\coords}},
    }
    \pgfplotstablenew[columns={level, coord}]{8}{\leveltable}
    
    \pgfplotstablegetelem{2}{coord}\of{\leveltable}
    \pgfmathsetmacro{\levelA}{\pgfplotsretval}
    \pgfplotstablegetelem{4}{coord}\of{\leveltable}
    \pgfmathsetmacro{\levelB}{\pgfplotsretval}
    \pgfplotstablegetelem{6}{coord}\of{\leveltable}
    \pgfmathsetmacro{\levelC}{\pgfplotsretval}
    
    \begin{axis}[
        name=main plot,
        width=0.75\columnwidth,
        height=0.75\columnwidth,
        colormap name=viridis,
        xlabel=Lower eigenvalue bound $\ubar{\lambda}$,
        ylabel=Upper eigenvalue bound $\bar{\lambda}$,
        xmin=-0.8,
        xmax=14,
        ymin=-1,
        ymax=21,
    ]
        \addplot[thin, domain=-1:15]{x}  node[pos=0.7, below, sloped]{$\ubar{\lambda} = \bar{\lambda}$};
        \addplot[thin, domain=-1:15]{20} node[pos=0.2, below]{$\bar{\lambda} = N$};
        
        \addplot[dashed] coordinates {(5,5) (5,20)};
        \coordinate (contourA) at (axis cs:\levelA+0.6,\levelA);
        \coordinate (contourB) at (axis cs:\levelB,\levelB);
        \coordinate (contourC) at (axis cs:14,\levelC);
        
       	\addplot[
            contour prepared,
            contour prepared format=matlab,
            contour/label distance=80pt,
            contour/contour label style={
                nodes={
                    text=black,
                    inner sep=1.1pt
                },
            },
        ] table {\contourtable};
    \end{axis}

    \begin{axis}[
        width=0.4\columnwidth,
        height=0.75\columnwidth,
        at={(main plot.north east)},
        xshift=0.3cm,
        anchor=north west,
        xlabel=$H_2$ bound $\gamma$,
        ytick=\empty,
        ymin=-1,
        ymax=21,
    ]
        \addplot table[col sep=comma, y=y_slc, x=h2_slc] {figures/data/eigenvalue_slice.csv};
        
        \coordinate (sliceA) at (axis cs:40,\levelA);
        \coordinate (sliceB) at (axis cs:60,\levelB);
        \coordinate (sliceC) at (axis cs:80,\levelC);
        \draw[dotted, semithick] (sliceA) -- (axis cs:40,-1);
        \draw[dotted, semithick] (sliceB) -- (axis cs:60,-1);
        \draw[dotted, semithick] (sliceC) -- (axis cs:80,-1);
        
    \end{axis}
    
    \draw[dotted, semithick] (contourA) -- (sliceA);
    \draw[dotted, semithick] (contourB) -- (sliceB);
    \draw[dotted, semithick] (contourC) -- (sliceC);
    
\end{tikzpicture}
    \caption{
        Contour plot of the best upper bound on the \(H_2\)-norm that can be obtained from Theorem~\ref{thm:h2_mas} for the consensus example.
        The second plot shows a slice of the upper bound on the \(H_2\)-norm for fixed \(\ubar{\lambda} = 5\), indicated by the dashed line in the contour plot.
    }
    \label{fig:eigenvalue_contour}
\end{figure}

The first effect we study is the influence of the size of the interval \([\ubar{\lambda}, \bar{\lambda}]\) on the \(H_2\)-norm bound.
Because of Assumption~\ref{ass:bounded_laplacian}, it is sufficient to check intervals adhering to the constraint \(0 < \ubar{\lambda} \leq \bar{\lambda} \leq N\).
Fixing the transmission probability to \(p = 0.5\), we obtain for each admissible interval the minimal \(\gamma\) such that there exist \(Q\) and \(Z_1\) satisfying the conditions in Theorem~\ref{thm:h2_mas}.
The contour lines of \(\gamma\) as a function of \(\ubar{\lambda}\) and \(\bar{\lambda}\) are shown in the left part of Figure~\ref{fig:eigenvalue_contour}.
In addition, the second part of the figure shows a slice of the function for constant \(\ubar{\lambda} = 5\) and \(5 \leq \bar{\lambda} \leq 20\).

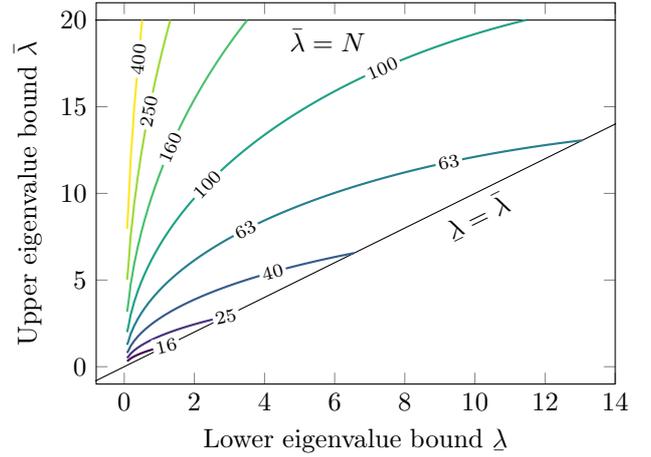
\begin{figure}
    \centering
    \begin{tikzpicture}
    \pgfplotsset{
        contour/label node code/.code={
            \node {
                \pgfkeys{/pgf/fpu=true}%
                \pgfmathparse{10^(#1)}%
                \pgfmathprintnumber[fixed relative]{\pgfmathresult}%
                \pgfkeys{/pgf/fpu=false}%
            };
        },
    }
    
    \begin{axis}[
        width=0.95\columnwidth,
        height=0.75\columnwidth,
        colormap name=viridis,
        xlabel=Lower eigenvalue bound $\ubar{\lambda}$,
        ylabel=Upper eigenvalue bound $\bar{\lambda}$,
        xmin=-0.8,
        xmax=14,
        ymin=-1,
        ymax=21,
    ]
        \addplot[thin, domain=-1:15]{x}  node[pos=0.675, below, sloped]{$\ubar{\lambda} = \bar{\lambda}$};
        \addplot[thin, domain=-1:15]{20} node[pos=0.425, below]{$\bar{\lambda} = N$};
        
       	\addplot[
            contour prepared,
            contour prepared format=matlab,
            contour/label distance=80pt,
            contour/contour label style={
                nodes={
                    text=black,
                    inner sep=1.1pt
                },
            },
        ] table[col sep=comma] {figures/data/eigenvalue_contour_adj.csv};
    \end{axis}
    
\end{tikzpicture}
    \caption{
        Contour plot of the best upper bound on the \(H_2\)-norm that can be obtained from Theorem~\ref{thm:h2_mas} for the input/output-switched system.
    }
    \label{fig:eigenvalue_contour_adjoint}
\end{figure}

For the consensus problem, only the upper bound \(\bar{\lambda}\) is relevant to the \(H_2\)-norm bound.
Furthermore, as shown in the second part of the figure, the relationship of \(\gamma\) and \(\bar{\lambda}\) is approximately linear for this value of \(\kappa\).
This is, however, a feature of the particular example considered here.
Switching input and output of the system, i.e., \(B^d = C^p = 0\) and \(B^p = C^d = 1\), the upper bound is instead described by the contour lines in Figure~\ref{fig:eigenvalue_contour_adjoint}.
Enlarging the interval \([\ubar{\lambda}, \bar{\lambda}]\) does still increase the bound on the \(H_2\)-norm, the relation is however no longer independent of \(\ubar{\lambda}\) nor linear.

\subsection{Estimating Conservatism}\label{sec:demo_conservatism}
In the second numerical study, we estimate the conservatism of the conditions in Theorem~\ref{thm:h2_mas} for the consensus example.
This is done by using Monte-Carlo techniques to obtain a lower bound on the \(H_2\)-norm, which can then be compared to the upper bound from Theorem~\ref{thm:h2_mas}.

\begin{figure}
    \centering
\tikzstyle{agent} = [draw, circle, fill=gray!30, minimum size=2.1em, inner sep=0.1em]
\tikzstyle{link}  = [>=Latex]

\begin{tikzpicture}[scale=1.55]
    \draw[link] (120:1) arc (120:-120:1);
    \draw[link, dotted] (120:1) arc (120:165:1);
    \draw[link, dotted] (240:1) arc (240:195:1); 
    \draw[link, dashed] (0:1) arc (-100:-180:1);
    \draw[link, dashed] (0:1) arc ( 100: 180:1);
    \draw[link, dashdotted] (0:1) arc (-90:-150:1.7);
    \draw[link, dashdotted] (0:1) arc ( 90: 150:1.7);

	\draw (  0:1) node[agent](a1){$v_1$};
	\draw ( 40:1) node[agent](a2){$v_2$};
	\draw ( 80:1) node[agent](a3){$v_3$};
    \draw (120:1) node[agent](a4){$v_4$};
	\draw (240:1) node[agent](a5){$v_{18}$};
	\draw (280:1) node[agent](a6){$v_{19}$};
	\draw (320:1) node[agent](a7){$v_{20}$};
\end{tikzpicture}

\DeclareRobustCommand{\drawlink}[1]{\tikz[baseline]{\draw[link, #1] (0,0.5ex) -- (0.6,0.5ex);}}
    \caption{
        Undirected circular graph with 20 vertices and one (\drawlink{}), two (\drawlink{dashed}), or three (\drawlink{dashdotted}) forward links.
    }
    \label{fig:circular_graphs}
\end{figure}
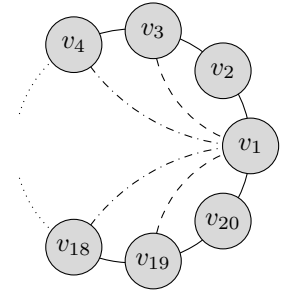

For this scenario, the interconnection structure is switched between one of seven circular graphs such that every agent has between one and seven forward and -- due to graph symmetry -- an equal number of backward neighbours.
The graph structure is visualized in Figure~\ref{fig:circular_graphs} for one up to three forward neighbours.
The eigenvalue bounds for this set of graphs are \(\ubar{\lambda} = 2.68\) and \(\bar{\lambda} = 18.24\).

\begin{figure}
    \centering
    \begin{tikzpicture}
    \pgfplotstableread[col sep=comma]{figures/data/performance_span.csv} \datatable

    \begin{axis}[
        width=0.95\columnwidth,
        height=0.75\columnwidth,
        xlabel=Transmission Probability $p$,
        ylabel=$H_2$-Performance,
        xmin=0,
        xmax=1,
        ymax=270,
    ]
        \addplot table[x=p_swp, y=h2_rb] {\datatable};
        
        \addplot[name path=min, black!60, forget plot] table[x=p_swp, y=h2_min] {\datatable};
        \addplot[name path=max, black!60, forget plot] table[x=p_swp, y=h2_max] {\datatable};
        
        \addplot[area legend, draw opacity=0.6, fill=gray!40] fill between[of=min and max];
        
        \legend{Bound from Theorem~\ref{thm:h2_mas}, Monte-Carlo estimates}
    \end{axis}
    
\end{tikzpicture}
    \caption{
        \(H_2\)-Performance of the first-order consensus system with different transmission probabilities.
        An upper bound is obtained from Theorem~\ref{thm:h2_mas} and Monte-Carlo simulations are used to estimate the norm for different switching sequences~\(\nu\), indicated by the gray band.
    }
    \label{fig:performance_span}
\end{figure}
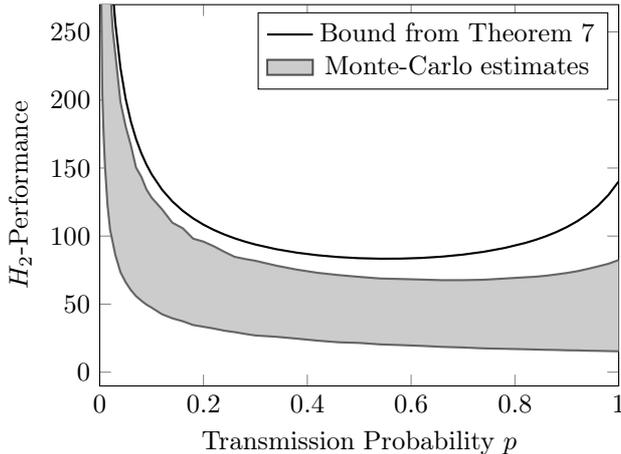

To accurately assess the conservatism of Theorem~\ref{thm:h2_mas}, it is necessary to find the worst-case switching sequence~\(\nu\) on the given set of graphs.
We will instead resort to finding a lower bound on the \(H_2\)-norm by evaluating the performance for a variety of switching sequences, amongst which are the constant sequences for each of the seven graphs, sequential switching between the graphs, and random switching.
This procedure is repeated for \(p \in (0, 1]\).
The span of performance measures obtained in this way is shown together with the upper bound provided by Theorem~\ref{thm:h2_mas} in Figure~\ref{fig:performance_span}.
For each switching sequence and probability, 10 random samples were drawn for the packet loss to evaluate the \(H_2\)-norm empirically.

The figure shows that the upper bound approximates the worst-case lower bound obtained from Monte-Carlo simulations reasonably well.
A significant widening of the gap is only observed for probabilities~\(p\) close to 1.
For \(p \to 0\), both bounds exhibit a steep decrease in performance.
This is consistent with \(p = 0\) being the critical probability of the system at which the disagreement dynamics are no longer stable (cf.  \citet{Schenato2007}).

\section{Conclusions}\label{sec:conclusions}
In this paper, we studied the system performance of MASs with time-varying interconnection structure and stochastic packet loss in terms of the \(H_2\)-norm.
In particular, LMI conditions that are of constant complexity independent of the number of agents and bound the \(H_2\)-norm from above were derived.
Finally, the conservatism of the LMI conditions was estimated for a first-order consensus example using Monte-Carlo simulations.

\bibliography{arxiv}

\appendix
\section{Proof of Theorem~\MakeLowercase{\ref{thm:h2}}}\label{app:h2_proof}
For the proof of Theorem~\ref{thm:h2}, we require the following lemma on bounding stochastic quadratic forms:
\begin{lem}\label{lem:stochastic_quadratic_form}
    Let \(v \in \real^n\) and \(M \in \real^{n \times n}\) be a random vector and a symmetric random matrix, respectively.
    Assume furthermore that \(v\) and \(M\) are independent random variables.
    Then for symmetric \(W\) with \(\expect[M] \preceq W\) it follows that
    \begin{equation*}
        \expect\big[v^T M v\big] \leq \expect\big[v^T W v\big].
    \end{equation*}
\end{lem}
\begin{pf}
    For symmetric matrices \(A\) and \(W\) with \(A \preceq W\) and an arbitrary \(B \succeq 0\) we have
    \begin{align*}
        A \preceq W
        &\quad\Rightarrow\quad B^{\frac 12} A B^{\frac 12} \preceq B^{\frac 12} W B^{\frac 12} \\
        &\quad\Rightarrow\quad \trace\big(B^{\frac 12} A B^{\frac 12}\big) \leq \trace\big(B^{\frac 12} W B^{\frac 12}\big) \\
        &\quad\Leftrightarrow\quad \trace(AB) \leq \trace(WB).
    \end{align*}
    From independence of \(v\) and \(M\) as well as linearity of expectation and trace it follows that
    \begin{alignat*}{3}
        \expect\big[v^T M v\big]
        &= \expect\big[\trace (v^T M v)\big] &
        &= \expect\big[\trace (M v v^T)\big] \\
        &= \trace\big(\expect[M v v^T]\big) &
        &= \trace\big(\expect[M] \expect[v v^T]\big)
    \end{alignat*}
    and taking \(A = \expect[M]\) and \(B = \expect[v v^T]\) we have
    \begin{alignat*}{3}
        \trace\big(\expect[M] \expect[v v^T]\big)
        &\leq \trace\big(W \expect[v v^T]\big) &
        &= \trace\big(\expect[W v v^T]\big) \\
        &= \expect\big[\trace (W v v^T)\big] &
        &= \expect\big[v^T W v\big],
    \end{alignat*}
    completing the claim above.
\end{pf}

Then, we are ready to proof Theorem~\ref{thm:h2}.
The approach for the proof was inspired by \citet{Duan2013}.
\begin{pf}
    From the LMIs~\eqref{eq:h2_lmi}, it follows that for all \(j \in \mathcal{K}\) there exist an \(\varepsilon_j > 0\) such that
    \begin{align*}
        \expect\big[A_{\sigma_k, j}^T Q A_{\sigma_k, j} - Q + C_{\sigma_k, j}^T C_{\sigma_k, j} \,\big|\, \nu_k = j\big] &\preceq -\varepsilon_j I \\
        \expect\big[B_{\sigma_k, j}^T Q B_{\sigma_k, j} - Z + D_{\sigma_k, j}^T D_{\sigma_k, j} \,\big|\, \nu_k = j\big] &\preceq -\varepsilon_j I
    \end{align*}
    for all \(k \geq 0\).
    Thus, with \(\bar{\varepsilon} \coloneqq \min_{j \in \mathcal{K}} \varepsilon_j\), the shorthand \(A_k \coloneqq A_{\sigma_k, \nu_k}\) and similarly for \(B_k\), \(C_k\) and \(D_k\), we have
    \begin{subequations}\label{eq:h2_proof_lmi}\begin{align}
            \expect\big[A_k^T Q A_k - Q + C_k^T C_k \big] &\preceq -\bar{\varepsilon} I \prec 0 \label{eq:h2_proof_lmi_gramian}\\
            \expect\big[B_k^T Q B_k - Z + D_k^T D_k \big] &\preceq -\bar{\varepsilon} I \prec 0 \label{eq:h2_proof_lmi_trace}
    \end{align}\end{subequations}
    for all \(k \geq 0\) and all \(\nu\).
    Furthermore, \eqref{eq:h2_proof_lmi_gramian} implies 
    \begin{equation}\label{eq:h2_proof_lmi_lyap}
        \expect\big[A_k^T Q A_k - Q\big] \preceq -\bar{\varepsilon} I
    \end{equation}
    for all \(k \geq 0\) and \(\nu\) as well.
    
    Introduce the storage function \(V(\xi) = \xi^T Q \xi\) and its increment \(\Delta V_k \coloneqq V(\xi_{k+1}) - V(\xi_k)\).
    Then \(V(0) = 0\), \(V(\xi) > 0\) for all \(\xi \neq 0\) and \(V(\xi) \leq c \|\xi\|^2\) for all \(\xi\) with \(c \coloneqq \|Q\|\).
    Additionally, Lemma~\ref{lem:stochastic_quadratic_form} and \eqref{eq:h2_proof_lmi_lyap} imply that
    \begin{equation*}
        \expect\big[\Delta V_k\big]
        = \expect\big[\xi_k^T (A_k^T Q A_k - Q) \xi_k\big]
        \leq -\bar{\varepsilon} \expect\big[\|\xi_k\|^2\big]
    \end{equation*}
    because \(\xi_k\) and \(\sigma_k\) are independent.
    Consider then any \(T \geq 0\) and \(\xi_0\) for the telescoping sum
    \begin{equation*}
        \sum_{k = 0}^T \expect\big[\Delta V_k\big]
        = \expect\big[V(\xi_{T+1})\big] - V(\xi_0)
        \leq -\bar{\varepsilon} \sum_{k = 0}^T \expect\big[\|\xi_k\|^2\big],
    \end{equation*}
    resulting in
    \begin{equation*}
        \expect\big[\|\xi_T\|^2\big]
        \leq \sum_{k = 0}^T \expect\big[\|\xi_k\|^2\big]
        \leq \frac{1}{\bar{\varepsilon}} V(\xi_0)
        \leq \frac{c}{\bar{\varepsilon}} \|\xi_0\|^2
    \end{equation*}
    for all \(T \geq 0\) and thus boundedness.
    In the limit, we have
    \begin{equation*}
        \lim_{T \to \infty} \sum_{k = 0}^T \expect\big[\|\xi_k\|^2\big] \leq \frac{c}{\bar{\varepsilon}} \|\xi_0\|^2 < \infty,
    \end{equation*}
    which implies \(\lim_{k \to \infty} \expect[\|\xi_k\|^2] = 0\) and thus mean-square stability.
    
    For bounding the \(H_2\)-norm, let \(\xi^s\) denote the stochastic state-sequence corresponding to the output \(z^{\nu,s}\) for some switching sequence~\(\nu\) and take \(B_k^s\), \(D_k^s\) as the \(s\)th column of \(B_k\) and \(D_k\), respectively.
    We then have \(\xi_0^s = 0\), \(\xi_1^s = B_0^s\) and \(\xi_{k+1}^s = A_k \xi_k^s\) for all \(k \geq 1\).
    Taking \(\Delta V_k^s\) as the storage function increment corresponding to \(\xi^s\), we get
    \begin{equation*}
        \sum_{k = 0}^T E[\Delta V_k^s] = \expect\big[V(\xi_{T+1}^s)\big] - V(\xi_0^s) = \expect\big[V(\xi_{T+1}^s)\big].
    \end{equation*}
    By stability, it follows that \(\lim_{T \to \infty} \sum_{k = 0}^T E[\Delta V_k^s] = 0\).
    On the other hand, the above values for \(\xi^s\) give
    \begin{align*}
        \sum_{k = 0}^\infty& E[\Delta V_k^s] = \expect\big[B_0^{sT} Q B_0^s\big] + \sum_{k = 1}^\infty \expect\big[\xi_k^{sT} (A_k^T Q A_k - Q) \xi_k^s\big].
    \end{align*}
    The mean output energy for a specific \(\nu\) is then
    \begin{align*}
        \|z^{\nu, s}\|^2 =& \sum_{k = 0}^\infty \expect\big[(z^{\nu,s})^T z^{\nu,s}\big] \\
        =& \expect\big[D_0^{sT} D_0^s\big] + \sum_{k = 1}^\infty \expect\big[\xi_k^{sT} C_k^T C_k \xi_k^s\big] \\
        =& \expect\big[B_0^{sT} Q B_0^s + D_0^{sT} D_0^s\big] \\
        &+ \sum_{k = 1}^\infty \expect\big[\xi_k^{sT} (A_k^T Q A_k - Q + C_k^T C_k) \xi_k^s\big] \\
        \leq& \expect\big[B_0^{sT} Q B_0^s + D_0^{sT} D_0^s\big],
    \end{align*}
    where in the third equality we added \(0 = \sum_{k = 0}^\infty E[\Delta V_k^s]\) and the inequality follows from Lemma~\ref{lem:stochastic_quadratic_form} and \eqref{eq:h2_proof_lmi_gramian} because \(\xi_k^s\) and \(\sigma_k\) are independent.
    The \(H_2\)-norm is then obtained as
    \begin{align*}
        \|P\|_{H_2}^2 =& \sup_{\nu} \sum_{s = 1}^{N n_w} \big\|z^{\nu,s}\big\|^2 \\
        \leq& \max_{\nu_0} \sum_{s = 1}^{N n_w} \expect\big[B_0^{sT} Q B_0^s + D_0^{sT} D_0^s\big] \\
        =& \max_{\nu_0} \trace\big( \expect[B_0^T Q B_0 + D_0^T D_0] \big)
        < \trace(Z)
        < \gamma ^2,
    \end{align*}
    where the second to last inequality follows from \eqref{eq:h2_proof_lmi_trace}.
\end{pf}

\section{Proof of Theorem~\MakeLowercase{\ref{thm:h2_mas}}}\label{app:h2_mas_proof}
\begin{pf}
    Start by specializing the LMIs~\eqref{eq:h2_lmi} for the decomposable MJLS~\eqref{eq:jls}.
    Inserting the model matrix structure~\eqref{eq:matrix_structure} and assuming \(Q = I_N \otimes \tilde{Q}\), we obtain
    \begin{align*}
        I_N &\otimes (B^{dT} \tilde{Q} B^d + D^{dT} D^{d})
        + L_j^2 \otimes (B^{pT} \tilde{Q} B^p + D^{pT} D^p) \\
        &+ L_j \otimes (B^{dT} \tilde{Q} B^p + B^{pT} \tilde{Q} B^d + D^{dT} D^p + D^{pT} D^d) \\
        &+ \expect\big[\tilde{L}_{\sigma_k} \big| \nu_k = j\big] \otimes (B^{dT} \tilde{Q} B^c + D^{dT} D^c) \\
        &+ \expect\big[\tilde{L}_{\sigma_k}^T \big| \nu_k = j\big] \otimes (B^{cT} \tilde{Q} B^d + D^{cT} D^d) \\
        &+ \expect\big[\tilde{L}_{\sigma_k}^T \tilde{L}_{\sigma_k} \big| \nu_k = j\big] \otimes (B^{cT} \tilde{Q} B^c + D^{cT} D^c) \\
        &+ \big(L_j \expect\big[\tilde{L}_{\sigma_k} \big| \nu_k = j\big]\big) \otimes (B^{pT} \tilde{Q} B^c+ D^{pT} D^c) \\
        &+ \big(\expect\big[\tilde{L}_{\sigma_k}^T \big| \nu_k = j\big] L_j\big) \otimes (B^{cT} \tilde{Q} B^p + D^{cT} D^p)
        \prec Z
    \end{align*}
    for \eqref{eq:h2_lmi_trace}.
    By Lemma~\ref{lem:expected_laplacians}, this is equivalent to
    \begin{align*}
        I_N &\otimes \big(B^{dT} \tilde{Q} B^d + D^{dT} D^{d}\big) \\
        +& L_j \otimes \big[B^{dT} \tilde{Q} B^p + B^{pT} \tilde{Q} B^d + D^{dT} D^p + D^{pT} D^d \\
        &\quad + p (B^{dT} \tilde{Q} B^c + B^{cT} \tilde{Q} B^d + D^{dT} D^c + D^{cT} D^d) \\
        &\quad + 2p(1-p) (B^{cT} \tilde{Q} B^c + D^{cT} D^c)\big] \\
        +& L_j^2 \otimes\big[B^{pT} \tilde{Q} B^p + D^{pT} D^p + p^2 (B^{cT} \tilde{Q} B^c + D^{cT} D^c) \\
        &\quad + p (B^{cT} \tilde{Q} B^p + B^{pT} \tilde{Q} B^c + D^{cT} D^p + D^{pT} D^c)\big]
        \prec Z
    \end{align*}
    and by algebraic matrix manipulation to
    \begin{equation}\label{eq:h2_mas_proof_lmi_asymptotic}\begin{aligned}
        \big(*\big)^T &\big(I_N \otimes \tilde{Q}\big) \big(I_N \otimes B^d + L_j \otimes (p B^c + B^p)\big) \\
        &+ \big(*\big)^T \big(I_N \otimes D^d + L_j \otimes (p D^c + D^p)\big) \\
        &+ 2p(1-p) L_j \otimes \big(B^{cT} \tilde{Q} B^c + D^{cT} D^c\big)
        \prec Z.
    \end{aligned}\end{equation}
    
    On the other hand, as shown in \citet[Lemma~9]{Hespe2022}, the LMIs~\eqref{eq:h2_lmi} are convex in \(\lambda\) such that finding common \(Q, Z_1\) at \(\ubar{\lambda}\) and \(\bar{\lambda}\) implies that the LMIs are satisfied on the whole interval \([\ubar{\lambda}, \bar{\lambda}]\).
    Therefore, \eqref{eq:h2_mas_lmi_trace} implies that for all diagonal \(\Lambda \in \real^{(N-1) \times (N-1)}\) with entries in \([\ubar{\lambda}, \bar{\lambda}]\) we have
    \begin{align*}
        (*)^T &\big(I_{N-1} \otimes Q\big) \big(I_{N-1} \otimes B^d + \Lambda \otimes (p B^c + B^p)\big) \\
        &+ (*)^T \big(I_{N-1} \otimes D^d + \Lambda \otimes (p D^c + D^p)\big)\\
        &+ 2p(1-p) \Lambda \otimes \big(B^{cT} Q B^{cT} + D^{cT} D^c\big) \prec I_{N-1} \otimes Z_1.
    \end{align*}
    Furthermore, notice that \eqref{eq:h2_mas_lmi_asymptotic_gramian} is \eqref{eq:h2_mas_lmi_gramian} with \(\lambda = 0\), thus with \(\bar{\Lambda} \coloneqq \diag(0, \Lambda)\) and \(\bar{Z} \coloneqq \diag(Z_2, I_{N-1} \otimes Z_1)\)
    \begin{equation}\label{eq:h2_mas_proof_lmi_decoupled}\begin{aligned}
        (*)^T &\big(I_N \otimes Q\big) \big(I_N \otimes B^d + \bar{\Lambda} \otimes (p B^c + B^p)\big) \\
        &+ (*)^T \big(I_N \otimes D^d + \bar{\Lambda} \otimes (p D^c + D^p)\big)\\
        &+ 2p(1-p) \bar{\Lambda} \otimes \big(B^{cT} Q B^{cT} + D^{cT} D^c\big) \prec \bar{Z}.
    \end{aligned}\end{equation}
    From Assumption~\ref{ass:bounded_laplacian}, it follows that for each \(L_j\), \(j \in \mathcal{K}\), there exists an orthogonal transformation~\(U\) such that \(U^T L_j U = \bar{\Lambda}\) for some \(\bar{\Lambda}\) as described above.
    Therefore, we can apply \(U \otimes I_{n_w}\) as a congruence transformation to \eqref{eq:h2_mas_proof_lmi_decoupled}, showing that \eqref{eq:h2_mas_proof_lmi_asymptotic} is satisfied for all \(j \in \mathcal{K}\).
    Analogous conclusions can be drawn for \eqref{eq:h2_lmi_gramian}.
    Finally, \(\trace(\bar{Z}) = \trace(Z_2) + (N-1) \trace(Z_1) < \beta^2 + (N-1) \gamma^2\).
\end{pf}

\end{document}